\documentstyle[12pt]{article}
\begin{document}
\begin{center}
INVERSE ANTICIPATING CHAOS SYNCHRONIZATION\\
E. M. Shahverdiev,\footnote{Permanent address: Institute of Physics, 370143 Baku,Azerbaijan} S.Sivaprakasam and K. A. Shore,\\
School of Informatics, University of Wales, Bangor, Dean Street, Bangor, LL57 1UT, Wales, UK\\
ABSTRACT\\
\end{center}
We report a new type of chaos synchronization:inverse anticipating synchronization, where a time delay chaotic system $x$ can drive another system $y$ in such a way that the driven system anticipates the driver by synchronizing with its inverse future state:$y(t)=-x(t+\tau)$,$\tau>0$. We extend the concept of inverse anticipating chaos synchronization to cascaded systems. We propose means for the experimental observation of inverse anticipating chaos synchronization in external cavity lasers.\\
PACS number(s):05.45.Xt, 05.45.Vx, 42.55.Px, 42.65.Sf\\
~\\
\indent Chaos synchronization [1] is of fundamental importance for various 
complex physical, chemical and biological systems [2]. Application of chaos 
synchronization can be found in secure communications, optimization of non-linear 
systems' performance, modeling brain activity and pattern recognition [2]. Synchronization of coupled 
chaotic systems restricts the evolution of synchronized systems to the synchronization manifold and therefore eliminates  some degrees of freedom of the joint system, and 
thus leading to significant reduction of complexity. In this context new types of chaos synchronization can be considered as a novel ways of reducing unpredictability  of chaotic dynamics. Time delay systems are ubiquitous in nature, technology and society due to 
finite signal transmission times, switching speeds and memory effects [3]. 
Therefore the study of chaos synchronization in these systems is of high 
practical importance. Because of their ability to generate high-dimensional chaos time delay systems are 
also good candidates for secure communications 
based on chaos synchronization [4]. In addition, time delay 
systems can be considered as a special case of spatio-temporal systems [5]. \\
\indent In this letter we report a new type of synchronization: inverse anticipating 
synchronization, where a time-delayed chaotic 
system $x$ drives another system $y$ in such a way that 
the driven system anticipates the driver by synchronizing with its inverse future 
state:$x(t)=-y(t-\tau)$ (in conditional representation:$x=-y_{\tau}$) or equivalently $y(t)=-x(t+\tau)$ with $\tau>0$. 
We investigate inverse anticipating synchronization    
between two coupled systems both with a single delay time and with 
two characteristic delay times where the delay time in the coupling is different from the delay time in the coupled systems themselves. We extend our findings to cascaded systems. We also propose means for the experimental 
observation of inverse anticipating synchronization phenomenon using external cavity laser diodes.\\
We define inverse anticipating synchronization as follows: the driver system 
$$\hspace*{6cm}\frac{dx}{dt}=-\alpha x + f(x_{\tau}),\hspace*{7.3cm}(1)$$
synchronizes with a driven system 
$$\hspace*{6cm}\frac{dy}{dt}=-\alpha y - f(x),\hspace*{7.5cm}(2)$$
on the inverse anticipating synchronization manifold 
$$\hspace*{7cm}x=-y_{\tau}. \hspace*{8.2cm}(3)$$
In order to obtain this result we introduce the error signal: $\Delta=x-(-y_{\tau})=x+y_{\tau}$. Then 
from eqs.(1) and (2) it follows that $\frac{d\Delta}{dt}=-\alpha \Delta$. 
In many representative cases, chaos 
synchronization can be understood from the existence of a global 
Lyapunov function of the error signals [6]. By analyzing the Lyapunov function $L=\frac{1}{2}\Delta^{2}$ we obtain that for $\alpha >0 $ the inverse anticipating 
synchronization manifold $x=-y_{\tau}$ is globally attracting and asymptotically stable. Throughout this paper to enhance the accessibility and practicality of our presentation, we confine ourselves to the demonstration of principles using specific examples from different areas of physics.\\
First we consider inverse anticipating synchronization in the following 
coupled Ikeda systems with a single delay time:
$$\hspace*{5cm}\frac{dx}{dt}=-\alpha x - \beta \sin x_{\tau};\frac{dy}{dt}=-\alpha y +\beta \sin x,\hspace*{4.3cm}(4)$$
where $\alpha >0, \beta >0$. Equations (4) play an important role in electronics and physiological 
studies [7]. Using the error dynamics approach given above one finds that 
$x=-y_{\tau}$ is the inverse anticipating chaos synchronization manifold.\\
Inverse anticipating chaos synchronization can also be found in systems with two characteristic delay times: a time delay $\tau_{1}$ in the coupled systems themselves 
and a coupling delay $\tau_{2}$ between the systems. In this case we find that the inverse anticipating time is $\tau_{1}-\tau_{2}$, ($\tau_{2}<\tau_{1}$). 
Consider the following unidirectionally coupled driver (x) and response (y) 
systems where feedback and coupling delays $\tau_{1}$ and $\tau_{2}$ are different. 
$$\hspace*{5cm}\frac{dx}{dt}=-\alpha x - \beta \sin x_{\tau_{1}};
\frac{dy}{dt}=-\alpha y + \beta \sin x_{\tau_{2}},\hspace*{4.3cm}(5)$$
where $\alpha>0$. We find that $x=-y_{\tau_{1}-\tau_{2}}$ is the inverse anticipating chaos synchronization manifold for (5), as the error $\Delta=x+y_{\tau_{1}-\tau_{2}}$ obeys the following dynamics:$\frac{d\Delta}{dt}=-\alpha \Delta$.\\
As another example with two characteristic delay times we consider the following 
modified version of the delay-coupled Ikeda model [8].
$$\hspace*{-5cm}\frac{dx}{dt}=-\alpha x + m_{1} \sin x_{\tau_{1}},$$
$$\hspace*{4cm}\frac{dy}{dt}=-\alpha y + m_{2} \sin y_{\tau_{1}} + m_{3}\sin x_{\tau_{2}},\hspace*{6.5cm}(6)$$
where $m_{1},m_{2}$ and $m_{3}$ are constants.\\
One finds that under the condition 
$$\hspace*{7cm}m_{2}=m_{1}+m_{3}.\hspace*{7cm}(7)$$
eqs.(6) also admits the inverse anticipating 
synchronization manifold is $x=-y_{\tau_{1} - \tau_{2}}$. It follows from the error 
$\Delta =x+ y_{\tau_{1} - \tau_{2}}$  dynamics:
$$\hspace*{5cm}\frac{d\Delta}{dt}=-\alpha\Delta + m_{2}\cos x_{\tau_{1}} \Delta_{\tau_{1}},\hspace*{6.5cm}(8)$$
The sufficient stability condition of the trivial solution $\Delta=0$ of eq.(8) can be found from 
Krasovskii-Lyapunov functional approach for the time delay systems [9]: $\alpha > \vert m_{2}\vert$. The condition (7) can be considered as the existence (necessary) condition for {\it inverse anticipating chaos synchronization} for the unidirectionally coupled modified Ikeda model. Notice that the analogous existence condition for {\it anticipating chaos synchronization} [7] $x=y_{\tau_{1} - \tau_{2}}$ is $m_{1}=m_{2}+m_{3}$. {\it Anticipating synchronization} for the modified delay-coupled Ikeda model was investigated {\it numerically} in [8]. Our analytical result on anticipating synchronization existence condition is in excellent agreement with numerics in [8]. Thus by changing the feedback and/or the coupling strengths one can make transitions between anticipating and inverse anticipating synchronizations and vice-versa. Such transitions, in principle can be exploited for the message encoding purposes.\\
\indent Studying the possibility of inverse anticipating 
synchronization in chaotic semiconductor lasers with optical feedback can be of importance, due to their ease of operation in high-speed optical communications and potential in secure communications [4]. External cavity laser diodes are commonly modeled with the Lang-Kobayashi equations, see, e.g. [8]. Here based on [10] we will use rate equations for the laser intensity $I$ and carrier density $n$. 
The use of the rate equations, which neglect the optical phase is justified in detail in [10]. (We  have also investigated inverse anticipating chaos synchronization regime in external cavity lasers modelled 
by the Lang-Kobayashi equations, which includes the optical phase and have derived the same existence condition for the inverse anticipating synchronization manifold as in the case of simpler rate equations presented here.)\\
Suppose that the master laser
$$\hspace*{-1.7cm}\frac{dI_{1}}{dt}=(gn_{1}-\gamma)I_{1}+k_{1}I_{1}(t-\tau_{1}),$$
$$\hspace*{5cm}\frac{dn_{1}}{dt}=a-\gamma_{e}n_{1}-gn_{1}I_{1},\hspace*{7.3cm}(9)$$
is coupled unidirectionally with the slave laser
$$\hspace*{0.3cm}\frac{dI_{2}}{dt}=(gn_{2}-\gamma)I_{2}+k_{2}I_{2}(t-\tau_{1}) + k_{3}I_{1}(t-\tau_{2}),$$
$$\hspace*{4.7cm}\frac{dn_{2}}{dt}=a-\gamma_{e}n_{2}-gn_{2}I_{2},\hspace*{7.8cm}(10)$$
where $g$ is the differential optical gain; $\tau_{1}$ is the external cavity round-trip time for the master and slave lasers ; $\tau_{2}$ is the coupling time between 
lasers; $\gamma_{e}$- the carrier decay rate; $\gamma$-the cavity decay 
rate (cavity losses); $a$ - the injection current; $k_{1}$, $k_{2}$ and $k_{3}$ are feedback and coupling rates, respectively. \\
We show that systems (9) and (10) allows for inverse anticipating chaos synchronization with $\tau_{1}> \tau_{2}$. First we use the standard procedure of change of dependent variables to eliminate the positive 
term $a$ in eqs.(9) and (10): $I_{1}=I_{1,s}+I_{1,x}$, $n_{1}=n_{1,s}+n_{1,x}$, 
$I_{2}=I_{2,s}+I_{2,x}$, $n_{2}=n_{2,s}+n_{2,x}$; where $I_{1,s}, n_{1,s}$ and 
$I_{2,s},n_{2,s}$ are the steady state solutions of systems (9) and (10),respectively. Next let $\Delta_{1}$ and $\Delta_{2}$ be error signals defined as  $\Delta_{1}=I_{1,x}-(-I_{2,x,\tau_{1}-\tau_{2}})=I_{1,x}+ I_{2,x,\tau_{1}-\tau_{2}}$ and 
$\Delta_{2}=n_{1,x}+n_{2,x,\tau_{1}-\tau_{2}}$. Then under the condition 
$$\hspace*{6cm}k_{2}=k_{1}+k_{3},\hspace*{8.5cm}(11)$$ 
it is possible to obtain the following decoupled error dynamics for $\Delta_{1}$ and
$\Delta_{2}$: 
$$\hspace*{4.5cm}\frac{d\Delta_{1}}{dt} + \gamma\Delta_{1} - k_{2}\Delta_{1,\tau_{1}}=-\frac{d\Delta_{2}}{dt} -\gamma_{e}\Delta_{2}.\hspace*{5cm}(12)$$
It is obvious that $\Delta_{1}=\Delta_{2}=0$ is solution of the system (12).
We underline that while deriving (12) we did not assume that the $\Delta_{1}$ 
and $\Delta_{2}$ are small. Notice that inverse anticipating chaos synchronization occurs between 
the dynamical variables in eqs.(9) and (10) after subtraction of corresponding steady state solutions.\\ 
The condition (11) is the necessary one for inverse anticipating chaos synchronization between unidirectionally coupled master and slave laser systems. This condition is analogous 
to the sum rule $k_{1}=k_{2}+k_{3}$ for feedback and coupling rates for the case of lag chaos synchronization 
$I_{2}=I_{1,\tau_{2}-\tau_{1}},n_{2}=n_{1,\tau_{2}-\tau_{1}}$,($\tau_{2}>\tau_{1}$) 
between unidirectionally coupled laser systems studied in [11]. Using eqs.(9) and (10) one also can obtain the necessary condition $k_{1}=k_{2}+k_{3}$
for anticipating chaos synchronization 
$I_{1}=I_{2,\tau_{1}-\tau_{2}},n_{1}=n_{2,\tau_{1}-\tau_{2}}$,($\tau_{1}>\tau_{2}$) 
between external cavity laser systems. This analytical result is confirmed by 
numerical simulations in [9]. From eqs.(9) and (10) it is also possible to derive the necessary (existence) condition $k_{2}=k_{1}+k_{3}$ for inverse lag chaos synchronization 
$I_{1,x}=-I_{2,x,\tau_{2}-\tau_{1}},n_{1,x}=-n_{2,x,\tau_{2}-\tau_{1}}$,($\tau_{2}>\tau_{1}$). 
Complete chaos synchronization  $I_{1}=I_{2}, n_{1}=n_{2}$,($\tau_{2}=\tau_{1}$) 
occurs, if $k_{1}=k_{2}+k_{3}$.  Inverse complete chaos synchronization
$I_{1,x}=-I_{2,x}, n_{1,x}=-n_{2,x}$,($\tau_{2}=\tau_{1}$) takes place, if $k_{2}=k_{1}+k_{3}$.
It is quite interesting that under existence conditions for all types of chaos synchronization corresponding errors dynamics always obey eq.(12) and therefore equation (12) has certain universality for unidirectionally coupled external cavity laser systems.\\
Equation (12) has a simple geometrical interpretation which is very helpful to find the stability condition of the inverse anticipating synchronization manifold: trivial solutions $\Delta_{1}=\Delta_{2}=0$ are the intersection points of a family of curves 
$\frac{d\Delta_{1}}{dt} + \gamma\Delta_{1} - k_{2}\Delta_{1,\tau_{1}}=c_{1}$
and $ -\frac{d\Delta_{2}}{dt} -\gamma_{e}\Delta_{2}=c_{2}$. As physically $\gamma_{e}$ is positive, we effectively arrive at the investigation of the stability of the trivial solution of equation $\frac{d\Delta_{1}}{dt} + \gamma\Delta_{1} - k_{2}\Delta_{1,\tau_{1}}=0$. An Krasovskii-Lyapunov functional approach allows to find sufficient stability condition: $\gamma > \vert k_{2}\vert$.\\
\indent Finally we demonstrate that the concept of cascaded synchronization allows one to obtain increased anticipation times using inverse 
anticipating chaos synchronization phenomenon. Consider the situation when the driven system in eq.(4) is a chain of, e.g. three response systems $y$, $z$ and $u$:
$\frac{dx}{dt}=-\alpha x - \beta \sin x_{\tau}$;$\frac{dy}{dt}=-\alpha y +\beta \sin x$;$\frac{dz}{dt}=-\alpha z - \beta \sin y$;$\frac{du}{dt}=-\alpha u - \beta \sin z$.
Then one can obtain that the driven system $z$ synchronizes with the driver 
system $x$ with the anticipation time $2\tau$: $x=-z_{2\tau}$. It follows from 
the investigation of the error dynamics $\Delta= x-(-z_{2\tau})=x+z_{2\tau}$:
$\frac{d\Delta}{dt}=-\alpha \Delta -\beta (\sin x_{\tau}+ \sin y_{2\tau})$. 
Assume that inverse anticipiating synchronization between $x$ and $y$ state variables has already taken place:$x=-y_{\tau}$; then with $x_{\tau}=-y_{2\tau}$  we arrive at the error dynamics:$\frac{d\Delta}{dt}=-\alpha \Delta$. Thus by adding new driven system to (4) it is possible to double the inverse anticipation time. In the case of three driven systems it is straightforward to find that $x=-u_{3\tau}$ is the inverse anticipating synchroniation manifold with the anticipation time $3\tau$.  It is straightforward to obtain that cascaded inverse anticipating synchronization allows to increase the anticipating times in the case of coupled systems (5) with two characteristic delays, too and thereby providing large anticipating times in a wide class of non-linear systems, including chaotic external cavity lasers.\\ 
\indent To summarize, we have reported a new type of chaos synchronization: inverse anticipating 
synchronization, where time delay chaotic system can drive another system in such a way that the driven system anticipates the driver by synchronizing with its inverse future state. This new type of chaos synchronization can be considered as a novel way of reducing unpredictability of chaotic dynamics. In order to exploit this phenomenon's capability of {\it anticipating inverse future states} of the master system, it is of great importance to obtain increased anticipating times. We have demonstrated here that the concept of cascaded slave systems can provide large anticipating times. Possible application areas of inverse anticipating chaos synchronization are: fast prediction, secure communications, controlling delay-induced instabilites in a wide range of non-linear systems. We have also found that the transition between anticipating and inverse anticipating 
chaos synchronizations can be realised by changing feedback and/or coupling rates.
We conjecture that this transition can be used for binary message 
encoding. In practice the efficiency of this method will be limited by the length of the transition period.\\
This work is supported by UK EPSRC under grants GR/R22568/01 and GR/N63093/01.\\

\end{document}